\documentclass[compsoc, conference, a4paper, 10pt, times]{IEEEtran}

\usepackage{cite}
\usepackage{amsmath,amssymb,amsfonts}
\usepackage{algorithmic}
\usepackage{graphicx}
\usepackage{textcomp}
\usepackage{xcolor}
\usepackage{booktabs}
\usepackage[hidelinks]{hyperref}
\usepackage{hhline}
\usepackage{colortbl}
\usepackage{multirow}
\usepackage{balance}
\usepackage{xcolor}
\usepackage{soul}
\usepackage{scalerel}
\usepackage{tikz}

\newcommand*\emptycirc[1][1ex]{\tikz\draw (0,0) circle (#1);} 
\newcommand*\halfcirc[1][1ex]{%
  \begin{tikzpicture}
  \draw[fill] (0,0)-- (90:#1) arc (90:270:#1) -- cycle ;
  \draw (0,0) circle (#1);
  \end{tikzpicture}}
\newcommand*\fullcirc[1][1ex]{\tikz\fill (0,0) circle (#1);}

\usepackage[strict]{changepage}
\usepackage{framed}
\definecolor{formalshade}{rgb}{0.85,1,0.85}
\definecolor{darkblue}{rgb}{0.0,0.6,0.30}
\newenvironment{formal}{%
  \MakeFramed{\advance\hsize-\width\FrameRestore}%
  \noindent\hspace{-4.55pt}%
  \begin{adjustwidth}{}{7pt}%
}
{%
  \end{adjustwidth}\endMakeFramed%
}

\usepackage[acronym, shortcuts]{glossaries-extra}

\glssetcategoryattribute{acronym}{nohyper}{true}
\setabbreviationstyle[acronym]{long-short}

\newacronym{acc}{ACC}{Adaptive Cruise Control}
\newacronym{ads}{ADS}{Anomaly Detection System}
\newacronym{ai}{AI}{Artificial Intelligence}
\newacronym{av}{AV}{Autonomous Vehicle}
\newacronym{bim}{BIM}{Basic Iterative Method}
\newacronym{dl}{DL}{Deep Learning}
\newacronym{dos}{DoS}{Denial of Service}
\newacronym{fgsm}{FGSM}{Fast Gradient Sign Method}
\newacronym{gps}{GPS}{Global Positioning System}
\newacronym{imu}{IMU}{Inertial Measurement Unit}
\newacronym{lidar}{LiDAR}{Light Detection And Ranging}
\newacronym{ml}{ML}{Machine Learning}
\newacronym{odd}{ODD}{Operational Design Domain}
\newacronym{sae}{SAE}{Society of Automotive Engineers}

\makeatletter
\def\ps@IEEEtitlepagestyle{%
  \def\@oddfoot{\mycopyrightnotice}%
  \def\@oddhead{\hbox{}\@IEEEheaderstyle\leftmark\hfil\thepage}\relax
  \def\@evenhead{\@IEEEheaderstyle\thepage\hfil\leftmark\hbox{}}\relax
  \def\@evenfoot{}%
}
\def\mycopyrightnotice{%
  \begin{minipage}{\textwidth}
  \centering \scriptsize
  \copyright~2024 IEEE. Personal use of this material is permitted. Permission from IEEE must be obtained for all other uses, in any current or future media, including reprinting/republishing this material for advertising or promotional purposes, creating new collective works, for resale or redistribution to servers or lists, or reuse of any copyrighted component of this work in other works by sending a request to pubs-permissions@ieee.org.
  \end{minipage}
}
\makeatother

\begin{document}

\title{Work-in-Progress:\\Crash Course: Can (Under Attack) Autonomous Driving Beat Human Drivers?}

\author{\IEEEauthorblockN{Francesco Marchiori}
\IEEEauthorblockA{\textit{University of Padua} \\
Padua, Italy \\
francesco.marchiori.4@phd.unipd.it}
\and
\IEEEauthorblockN{Alessandro Brighente}
\IEEEauthorblockA{\textit{University of Padua} \\
Padua, Italy \\
alessandro.brighente@unipd.it}
\and
\IEEEauthorblockN{Mauro Conti}
\IEEEauthorblockA{\textit{University of Padua} \\
Padua, Italy \\
\textit{Delft University of Technology} \\
Delft, Netherlands \\
mauro.conti@unipd.it}
}

\maketitle

\begin{abstract}
Autonomous driving is a research direction that has gained enormous traction in the last few years thanks to advancements in Artificial Intelligence (AI).
Depending on the level of independence from the human driver, several studies show that Autonomous Vehicles (AVs) can reduce the number of on-road crashes and decrease overall fuel emissions by improving efficiency.
However, security research on this topic is mixed and presents some gaps.
On one hand, these studies often neglect the intrinsic vulnerabilities of AI algorithms, which are known to compromise the security of these systems.
On the other, the most prevalent attacks towards AI rely on unrealistic assumptions, such as access to the model parameters or the training dataset.
As such, it is unclear if autonomous driving can still claim several advantages over human driving in real-world applications.

This paper evaluates the inherent risks in autonomous driving by examining the current landscape of AVs and establishing a pragmatic threat model.
Through our analysis, we develop specific claims highlighting the delicate balance between the advantages of AVs and potential security challenges in real-world scenarios.
Our evaluation serves as a foundation for providing essential takeaway messages, guiding both researchers and practitioners at various stages of the automation pipeline.
In doing so, we contribute valuable insights to advance the discourse on the security and viability of autonomous driving in real-world applications.
\end{abstract}

\section{Introduction}
\label{sec:introduction}

\acp{av}, also known as self-driving or driverless vehicles, refer to vehicles that do not need any form of human interaction to operate.
An \ac{av} is capable of sensing the environment and recognizing the route, the roadway, and possible obstacles, as well as managing the speed and acceleration of the car.
This is achieved through the cooperation of signals from various sensors, such as \ac{gps}, radars, \acp{lidar}, and thermographic cameras.
While the final stage of \acp{av} are still being developed and are not ready for mass production yet, the advantages that can come with them are clear: reduction of vehicle collisions, overall reduction of fuel consumption, increased comfort for passengers, and, most importantly, safer transport~\cite{yang2014vehicle}.
Given the complexity of such signals, data-driven approaches such as \ac{ml} and \ac{dl} must be used to elaborate all the information and decide the actions to perform in real time.
However, this increased reliance on the technological components of the vehicle and the need for communication with its surroundings poses some security concerns.
Indeed, while the cameras installed on the vehicle's various parts can implement a robust computer vision system, even a single error could precipitate collisions and crashes with far-reaching consequences.
Processing these signals is often handled by \ac{ml} or \ac{dl} models, which, however, are inherently vulnerable to adversarial attacks~\cite{goodfellow2014explaining}.
One example is \textit{evasion attacks}, in which the attacker can target the computer vision system to impose misclassification at test time.
This attack can be applied to street road signs, a scenario in which the consequences can be profound and pose significant risks to public safety and traffic management systems~\cite{morgulis2019fooling}.

\paragraph{\textit{Contribution}}
This paper delves into the question: \textit{can autonomous driving systems outperform human drivers when subjected to adversarial attacks?}
Traditionally, the focus has been on the efficiency and safety of \acp{av} in ideal conditions.
However, the emergence of evasion attacks~\cite{goodfellow2014explaining}, poisoning attacks~\cite{tian2022comprehensive}, and other adversarial techniques raise critical concerns about the robustness of autonomous driving algorithms.
This paper explores the concrete risks associated with autonomous driving under attack.
By defining a realistic threat model and conducting comparative analyses between autonomous and human drivers, we seek to shed light on the capabilities and vulnerabilities of \ac{ai} systems.
Furthermore, we scrutinize the current state-of-the-art to highlight and evaluate the assumptions on attacks towards \acp{av}.
Our findings have implications for the technological advancement of autonomous driving and the broader discourse on integrating \ac{ai} into safety-critical applications.

Our contributions can be summarized as follows.
\begin{itemize}
    \item We evaluate the vulnerabilities of all levels of automation for \acp{av}.
    We highlight possible threats in real-world scenarios by formalizing the contribution of \ac{ai} vs. human drivers in each of these levels.
    \item We delineate a realistic threat model in the context of attacks to \acp{av}.
    We identify differences between common assumptions in adversarial attacks and realistic knowledge of a possible attacker.
    \item We perform a literature review on attacks to \acp{av} and discuss their requirements to be successful.
    We identify four common requirements and discuss their feasibility in real-world scenarios.
    \item We provide a list of security requirements and suggestions for safe and secure implementation of \ac{ai} systems in \acp{av}.
\end{itemize}
\paragraph{\textit{Organization}}
The paper is organized as follows.
In Section~\ref{sec:background}, we give some background on \acp{av}.
Section~\ref{sec:related} delves into the current literature and highlights the assumptions of related works on attacks towards \acp{av}.
Section~\ref{sec:threatmodel} proposes a realistic threat model in the current landscape based on this discussion.
In Section~\ref{sec:criteria}, we identify the criteria for our evaluation, and in Section~\ref{sec:claims}, we analyze the vulnerabilities of each level of automation.
In Section~\ref{sec:takeaways}, we distill the takeaway messages from our study, and Section~\ref{sec:conclusions} concludes this work.
\section{Background}
\label{sec:background}

The \ac{sae} defines six levels of automation~\cite{sae, iso22736}: starting from level 0, where the vehicle only provides warning messages and momentary assistance, we reach level 5, where the human driver is not needed in any road condition.
We show an overview in Table~\ref{tab:levels}.
We can divide \ac{sae} levels into two significant groups. 
In the first, the vehicle provides driver support features such as automatic emergency braking (level 0), lane centering (level 1), and both lane centering and adaptive cruise control (level 2). 
In the second group, the vehicle is equipped with automated driving features such as traffic jam chauffeur (level 3), local driverless taxis in geofenced areas (level 4), and driverless vehicles in all regions (level 5). 
Furthermore, with levels 4 and 5, pedals and steering wheels may not be installed, preventing any traditional control from the driver to the vehicle.

\begin{table*}[!t]
\centering
\caption{\ac{sae} Levels of Automation and Involvement of \ac{ai}.}
\label{tab:levels}
\begin{tabular}{c|c|l|c|c|l}
\hline
\textbf{Level} & \textbf{Automation} & \textbf{Example Features} & \textbf{AI} & \textbf{Driver} & \textbf{Example Tasks} \\ \hline
\rowcolor{gray!15}
0 & - & - & \emptycirc & \fullcirc & - \\ \hline

 &  & \ac{acc} & \fullcirc & \fullcirc & Decision making \\
\multirow{-2}{*}{1} & \multirow{-2}{*}{\begin{tabular}[c]{@{}c@{}}Partial\\Assistance\end{tabular}} & Lane departure warning & \fullcirc & \fullcirc & Detection, sensor fusion \\ \hline

\rowcolor{gray!15} &  & \ac{acc} & \fullcirc & \fullcirc & Decision making \\ 
\rowcolor{gray!15} &  & Lane keeping assistance & \fullcirc & \fullcirc & Detection, sensor fusion \\
\rowcolor{gray!15} &  & Driver monitoring & \fullcirc & \fullcirc & Biometrics analysis \\
\rowcolor{gray!15} \multirow{-4}{*}{2} & \multirow{-4}{*}{\begin{tabular}[c]{@{}c@{}}Partial\\Automation\end{tabular}} & Traffic jam assistant & \fullcirc & \fullcirc & Traffic pattern recognition \\ \hline

 &  & Environment monitoring & \fullcirc & \emptycirc & Sensor fusion \\ 
 &  & Traffic jam autopilot & \fullcirc & \halfcirc & Autonomous decision making \\
 &  & Driver disengagement & \fullcirc & \halfcirc & Autonomous decision making \\ 
\multirow{-4}{*}{3} & \multirow{-4}{*}{\begin{tabular}[c]{@{}c@{}}Conditional\\Automation\end{tabular}} & Autonomous driving & \fullcirc & \halfcirc & Lane change, navigation \\ \hline

\rowcolor{gray!15} &  & Navigation in geofenced areas & \fullcirc & \emptycirc & Path planning \\
\rowcolor{gray!15} &  & Autonomous decision making & \fullcirc & \emptycirc & Traffic management \\
\rowcolor{gray!15} \multirow{-3}{*}{4} & \multirow{-3}{*}{\begin{tabular}[c]{@{}c@{}}High\\Automation\end{tabular}} & Safety overrides & \fullcirc & \halfcirc & Limited safety-critical tasks \\ \hline

 &  & Safety and redundancy & \fullcirc & \emptycirc & Anomaly detection \\
 &  & V2X communications & \fullcirc & \emptycirc & Resource optimization \\
\multirow{-3}{*}{5} & \multirow{-3}{*}{\begin{tabular}[c]{@{}c@{}}Full\\Automation\end{tabular}} & Navigation & \fullcirc & \emptycirc & Autonomous navigation \\ \hline
 
\multicolumn{6}{l}{\footnotesize{\fullcirc: present, \emptycirc: not present, \halfcirc: partially present.}}
\end{tabular}
\end{table*}

To achieve these capabilities, an autonomous driving system includes the sensing, perception, planning, and actuation layers~\cite{kim2022drivefuzz}. 
The sensing layer leverages technologies such as \ac{lidar}, cameras, radar, \ac{gps}, and \ac{imu} to acquire information from the surrounding environment.
The perception layer then leverages the collected data by fusing and interpreting them to comprehend the vehicle's surroundings. 
The planning layer then leverages the perceived state to design a routing plan for the vehicle. 
The general approach is to generate a global trajectory with intermediate waypoints and adjust it in real-time to reach the destination safely. 
Lastly, the actuation layer defines a concrete motion plan, including acceleration and deceleration patterns and steering wheel angles.

Most of the applications of \ac{ml} in autonomous driving are related to the perception and planning layers \cite{zhao2023autonomous,peng2020first}. 
The images collected via cameras need to be processed to detect lanes, traffic signs, and traffic lights and interpret them. 
Furthermore, \ac{ml} plays a fundamental role in detecting pedestrians and other moving objects that may impact road safety. Sensor fusion algorithms leverage \ac{ml} to extract meaningful information from different sensing sources. 
The planning layer makes an even more extensive use of \ac{ml}.
Indeed, algorithms for path searching, path planning, cooperative decision-making, collision avoidance, and decision generation all leverage \ac{ml} mostly with deep, recurrent, and generative structures \cite{peng2020first}. 
Lastly, \ac{ml} has mainly been employed to develop intrusion and anomaly detection at the previously described layers. 
\section{Related Works}
\label{sec:related}

In the literature, it is possible to find different kinds of adversarial attacks able to fool almost any \ac{ml} model.
The two most prevalent and effective attacks are \textit{evasion attacks} and \textit{poisoning attacks}.
Evasion attacks consist of carefully crafted noise to apply to samples at test time to drive the model to misclassification~\cite{goodfellow2014explaining}.
The noise is created by accessing the model parameters and is scaled to avoid human detection.
Poisoning attacks involve injecting malicious data into the training set to manipulate model behavior during testing~\cite{tian2022comprehensive}.
By tampering with training samples, attackers aim to influence the model's decision boundaries, often with subtle alterations to evade detection by human observers.
However, in some adversarial experimental settings, datasets are too simple and fail to accurately represent realistic scenarios in which \ac{dl} models could be deployed.
Other studies propose attacks that can fool the state-of-the-art with high attack success rates but require extensive knowledge of the dataset or the model parameters to be effective.

Table~\ref{tab:sota} summarizes the current state-of-the-art attacks towards \acp{av} and the assumptions and requirements needed to succeed.
Here, we focus on papers on evasion attacks that focus on \ac{av} applications or mention and test their methodology on \ac{av} tasks.
We decide not to include poisoning attacks since this technique requires access to the training dataset.
Indeed, in real-world scenarios, automotive companies will be required to produce their datasets and manually validate them to avoid those threats.
We identify four requirements and assumptions that are commonly made in the literature.
\begin{itemize}
    \item \textbf{Model Parameters} --
    When targeting models in white-box scenarios, it is common to use techniques such as \ac{fgsm}~\cite{goodfellow2014explaining} or \ac{bim}~\cite{kurakin2018adversarial} to craft adversarial samples.
    However, these techniques require access to the models' gradients as it is needed in their optimization process.
    As such, a real-world attacker aiming to compromise a \ac{dl} system with these methods would require complete access to the models, an assumption that is often unrealistic in many current scenarios.
    \item \textbf{Model Output} --
    Other adversarial attack techniques work in supposed black-box scenarios, i.e., do not require access to models' parameters to generate adversarial examples~\cite{ilyas2018black, andriushchenko2020square}.
    As such, the assumptions on the attacker's knowledge are far reduced, as they would require only access to the system's output.
    Indeed, several of these techniques query the target model in their adversarial sample generation process.
    In real-world scenarios, this assumption might require additional effort from the attacker, as they would need to own the target vehicle's image recognition model to be able to query it.
    \item \textbf{Direct Input} --
    The most common adversarial attacks toward image recognition systems use imperceptible perturbations computed through the model's gradient to cause misclassification.
    The adversarial noise is applied to each sample and then fed to the target model.
    However, applying these perturbations on physical objects, such as street signs, is particularly challenging, as it would require the complete re-print of the image.
    Thus, the only way to apply the adversarial noise at test time is to have control of the models' input.
    In this scenario, an attacker would need to modify the images captured by the cameras before they are fed to the model.
    While this assumption can be realistic if an attacker has gained control of the \ac{av} internal network (and thus carrying a man-in-the-middle attack), it relies on fundamental vulnerabilities of the vehicle bus.
    \item \textbf{Physical Implementation} --
    To solve the challenge posed by the \textit{direct input} assumption, several papers have delved into the feasibility of adversarial attacks using patches to apply to the target object~\cite{brown2017adversarial, eykholt2018robust}.
    While this requirement can be more realistic in most \ac{av} scenarios, the attacker must know the road on which the target vehicle will drive and carefully apply the patch, as environmental conditions can challenge the attack's success.
\end{itemize}

\begin{table*}[!htpb]
\centering
\caption{Requirements for state-of-the-art evasion attacks in \ac{av} tasks.}
\label{tab:sota}
\begin{tabular}{l|c|c|c|c|c}
\hline
\textbf{Attack} & \begin{tabular}[c]{@{}c@{}}\textbf{Misclassification}\\\textbf{Task}\end{tabular} & \begin{tabular}[c]{@{}c@{}}\textbf{Model}\\\textbf{Parameters}\end{tabular} & \begin{tabular}[c]{@{}c@{}}\textbf{Model}\\\textbf{Output}\end{tabular} & \begin{tabular}[c]{@{}c@{}}\textbf{Direct}\\\textbf{Input}\end{tabular} & \begin{tabular}[c]{@{}c@{}}\textbf{Physical}\\\textbf{Implementation}\end{tabular} \\ \hline

\rowcolor{gray!15}
Arnab et al.~\cite{arnab2018robustness} & Semantic Segmentation & \fullcirc & \fullcirc & \fullcirc & \emptycirc \\ \hline
Brown et al.~\cite{brown2017adversarial} & Road Sign & \fullcirc & \fullcirc & \emptycirc & \fullcirc \\ \hline
\rowcolor{gray!15}
Cao et al.~\cite{cao2019adversarial} & \ac{lidar} & \fullcirc & \fullcirc & \emptycirc & \fullcirc \\ \hline
Cao et al.~\cite{cao2019adversarial2} & \ac{lidar} & \emptycirc & \fullcirc & \emptycirc & \fullcirc \\ \hline
\rowcolor{gray!15}
Eykholt et al.~\cite{eykholt2018robust} & Road Sign & \fullcirc & \fullcirc & \emptycirc & \fullcirc \\ \hline
Kong et al.~\cite{kong2020physgan} & Road Sign & \emptycirc & \fullcirc & \emptycirc & \fullcirc \\ \hline
\rowcolor{gray!15}
Kumar et al.~\cite{kumar2020black} & Road Sign & \emptycirc & \fullcirc & \fullcirc & \emptycirc \\ \hline
Li et al.~\cite{li2020adaptive} & Road Sign & \emptycirc & \fullcirc & \fullcirc & \emptycirc \\ \hline
\rowcolor{gray!15}
Ma et al.~\cite{ma2023wip} & Object Tracking & \fullcirc & \fullcirc & \emptycirc & \fullcirc \\ \hline
Papernot et al.~\cite{papernot2017practical} & Road Sign & \emptycirc & \fullcirc & \fullcirc & \emptycirc \\ \hline
\rowcolor{gray!15}
Sharma et al.~\cite{sharma2019attacks} & Misbehavior Detection & \emptycirc & \fullcirc & \fullcirc & \emptycirc \\ \hline
Sitawarin et al.~\cite{sitawarin2018darts} & Road Sign & \emptycirc & \fullcirc & \fullcirc & \emptycirc \\ \hline
\rowcolor{gray!15}
Xiang et al.~\cite{xiang2019generating} & \ac{lidar} & \fullcirc & \fullcirc & \fullcirc & \emptycirc \\ \hline
Zhu et al.~\cite{zhu2021adversarial} & \ac{lidar} & \emptycirc & \fullcirc & \emptycirc & \fullcirc \\ \hline
\multicolumn{6}{l}{\footnotesize{\fullcirc: required, \emptycirc: not required.}}
\end{tabular}
\end{table*}
\section{Threat Model}
\label{sec:threatmodel}

Based on the related works discussed in Section~\ref{sec:related}, we identify four key factors that differentiate the current literature from real-world attackers.
\begin{itemize}
    \item \label{l1}
    \textbf{Limited Access to Model Architecture} --
     In many practical deployments, the detailed architecture of the autonomous driving model is proprietary and closely guarded by manufacturers or developers.
     Adversaries are unlikely to have unrestricted access to the intricate details of the model's structure, layers, and parameters.
    \item \label{l2}
    \textbf{Restricted Knowledge of the Training Data} -- 
    The datasets used to train autonomous driving models often consist of diverse real-world scenarios, making them extensive and complex.
    Realistically, adversaries would not possess exhaustive knowledge about the complete training dataset, limiting their ability to craft highly tailored attacks.
    \item \label{l3}
    \textbf{Constrained Sensor Data Manipulation} --
    Autonomous vehicles rely on sensors such as cameras, \ac{lidar}, and radar to perceive their environment.
    Manipulating these sensor inputs in real-time poses a substantial challenge, as physical access to the vehicle and the ability to tamper with sensor hardware is a great barrier for potential attackers.
    \item \label{l4}
    \textbf{Environmental Variability} --
    Real-world driving conditions encompass many environments, weather conditions, and traffic scenarios.
    Adversaries attempting to launch successful attacks must contend with the inherent variability in the operational environment, making it challenging to devise universally effective exploits.
\end{itemize}

In the case of evasion attacks, it is worth noting that they can be transferable, i.e., they can be computed on surrogate models to attack another model.
However, it has been shown that this solution is not optimal and is often outclassed by simple non-gradient-based manipulations of the input~\cite{alecci2023dumb}.
While poisoning and backdoor attacks also show a degree of transferability, they rely on access to the training dataset, which is limited~\cite{liu2023transferable}.
\section{Criteria}
\label{sec:criteria}

We now identify the criteria we use for discussing the security of each \ac{sae} level of automation.
Indeed, depending on which tasks are outsourced to \ac{ml} algorithms, four different properties determine the advantage of humans or \ac{ai} in specific scenarios.
The criteria are formulated through a synthesis of existing research in autonomous driving, machine learning, and adversarial attacks.
As such, they are grounded in a thorough examination of the pertinent literature and expert consensus within the domain.

\begin{itemize}
    \item \textbf{Ease of Attack} --
    This criterion assesses the simplicity or difficulty of executing an adversarial attack on human and \ac{ai} drivers.
    It considers factors such as the technical expertise required and the accessibility of attack methods.
    \item \textbf{Response Time} --
    Response time evaluates the speed and efficiency of human and \ac{ai} drivers react to adversarial situations.
    It reflects the ability to make quick decisions and take appropriate actions in response to unforeseen challenges.
    \item \textbf{Recovery Time} --
    Recovery time measures how swiftly human and \ac{ai} drivers can recover and resume normal operation following a successful adversarial attack.
    It determines the resilience and adaptability of each system to bounce back from disruptions.
    \item \textbf{Adaptability} --
    Adaptability examines how well human and \ac{ai} drivers can adjust and learn from adversarial encounters.
    It considers the capacity to evolve and improve responses, adapting to new attack strategies.
\end{itemize}
\section{Evaluation}
\label{sec:claims}

We now discuss and evaluate the levels of \ac{av} automation.
We divide our discussion into two parts.
First, we focus on each \ac{sae} automation level and discuss the feasibility of attack w.r.t. our threat model (Section~\ref{subsec:eval_threat}).
Then, we focus on our identified criteria and determine the advantages of \ac{ai} or human drivers for each of them (Section~\ref{subsec:eval_criteria}).

\subsection{Evaluation on Threat Model}
\label{subsec:eval_threat}

We now analyze each level of the \ac{sae} scale of automation, and by contextualizing them on the threat model, we provide insights into their advantages over human driving.

\paragraph{\textit{Level 1 - Partial Assistance}}
Despite the limited automation, \acp{av} at this level can provide valuable assistance in specific driving tasks, such as steering or acceleration (but not simultaneously).
As such, evasion attacks can occur.
However, the limitations of the threat model might not make these attacks feasible, given restricted access to the model architecture or training data.
Incremental safety improvements and potential fuel efficiency benefits suggest Level 1 \acp{av} may offer advantages over human drivers in specific scenarios but still rely on their judgment.

\paragraph{\textit{Level 2 - Partial Automation}}
Similarly to the previous level, \acp{av} in this category help to steer and accelerate but can do that simultaneously.
Thus, the same takeaways from the previous level apply, with different risks involved in the interaction between the assisted actions.
However, without access to the training data, it is unlikely to build hidden triggers in the model to cause harm.

\paragraph{\textit{Level 3 - Conditional Automation}}
\acp{av} at this level automate several driving phases but still require driver attention to take over if prompted.
As such, potential challenges might occur during the handover (i.e., the transfer of control from the automated system back to the human driver).
Indeed, as the number of automated processes increases, the attacker's limitations on sensor data manipulations become fewer since it grants the attacker multiple vectors for causing safety issues.
However, the limited access to the model architecture makes attack attempts easier to detect, and consequently, human drivers can take over.

\paragraph{\textit{Level 4 - High Automation}}
Since drivers may not need to intervene at this level, exploiting limitations beyond predefined scenarios is risky.
At this stage, it becomes imperative that the attacker cannot access the model parameters or tamper with the dataset.
If those measures are ensured, level 4 \acp{av} present enhanced safety and energy-efficient operation within specified contexts.

\paragraph{\textit{Level 5 - Full Automation}}
Comprehensive safety benefits, enhanced fuel efficiency, and increased accessibility suggest level 5 \acp{av} may outclass humans in a broader range of driving scenarios.
However, this is also the most challenging scenario, as it challenges system reliability and ethical decision-making.

\subsection{Evaluation on Criteria}
\label{subsec:eval_criteria}

Instead of focusing on each \ac{sae} automation level, we now highlight each criterion's impact at the varying number of automated components in \acp{av}.
A summary of our analysis is shown in Table~\ref{tab:eval}.

\paragraph{\textit{Ease of Attack}}
Assuming the attacker's knowledge is still limited according to our system model, attacking \acp{av} becomes more feasible as the level of automation increases.
Indeed, at level 1 or 2, malicious actors are limited to the restricted tasks assigned to \ac{ai} algorithms.
As such, attacks need to be specifically targeted towards specific models.
Given the increased number of automated components at higher automation levels, there is a higher chance of disrupting the systems with generic and black-box adversarial attacks.
Thus, while it might still be challenging to cause targeted misclassification, performing a \ac{dos} attack becomes easier.

\paragraph{\textit{Response Time}}
In unconstrained situations where the attacker has unlimited knowledge of the system, \acp{av} would become more vulnerable as the level of automation increases since attacks cannot be detected.
As such, they would rely on human take-over (available only in levels 1, 2, and 3) to promptly respond.
However, given the restricted assumptions allowed by our realistic threat model, attack attempts are not as successful as they would be.
Therefore, there is the possibility of implementing \acp{ads} to anticipate the presence of an attacker and thus react accordingly.
Once an attack is detected, human take-over can be prompted if present, or other safety measures can be employed to prevent its effect.

\paragraph{\textit{Recovery Time}}
If an attack is successful, an increased level of automation might be more effective for mitigating its effect.
The restricted automated capabilities of \acp{av} in the first levels prevent them from having sufficient aptitude and adaptability to the environment.
Therefore, it might be more challenging to resume normal operations, as external conditions might have drastically changed.
Instead, at high levels of automation, an \ac{av} is expected to operate in any environment, as human drivers are often not present in the system model.
They would thus be able to recover from the attack and mitigate it more promptly.

\paragraph{\textit{Adaptability}}
At lower levels of automation (1 and 2), the adaptability of \acp{av} is limited, relying heavily on human intervention for decision-making in complex scenarios.
In these situations, the adaptability of human drivers surpasses that of the automated systems. However, as automation levels progress, especially level 3 and beyond, the \ac{av}'s ability to adapt to various attack scenarios can increase.
Indeed, while there may be challenges during the handover phase in level 3, the overall adaptability rises, leveraging automated processes.
At higher levels (4 and 5), where driver intervention may not be needed, \acp{av} can exhibit advanced adaptability to diverse adversarial situations, continuously learning and improving responses over time.
This also relates to the response time considerations, as the constrained threat model challenges the attack effectiveness and thus allows the system to not only prevent them but also learn from them through techniques such as adversarial training.

\begin{table}[!htpb]
\centering
\caption{Contribution of \ac{ai} in \acp{av} for safety under different criteria for all \ac{sae} automation levels.}
\label{tab:eval}
\begin{tabular}{c|c|c|c|c}
\hline
\textbf{Level} & \begin{tabular}[c]{@{}c@{}}\textbf{Ease of}\\\textbf{Attack}\end{tabular} & \begin{tabular}[c]{@{}c@{}}\textbf{Response}\\\textbf{Time}\end{tabular} & \begin{tabular}[c]{@{}c@{}}\textbf{Recovery}\\\textbf{Time}\end{tabular} & \textbf{Adaptability} \\ \hline

\rowcolor{gray!15}
1 & \fullcirc & \halfcirc & \emptycirc & \emptycirc \\ \hline

2 & \fullcirc & \halfcirc & \emptycirc & \emptycirc \\ \hline

\rowcolor{gray!15}
3 & \halfcirc & \fullcirc & \halfcirc & \halfcirc \\ \hline

4 & \emptycirc & \fullcirc & \fullcirc & \fullcirc \\ \hline

\rowcolor{gray!15}
5 & \emptycirc & \fullcirc & \fullcirc & \fullcirc \\ \hline

\multicolumn{5}{l}{\footnotesize{\fullcirc: increased safety.}}\\
\multicolumn{5}{l}{\footnotesize{\halfcirc: unclear.}}\\
\multicolumn{5}{l}{\footnotesize{\emptycirc: no improvement or decreased safety.}}
\end{tabular}
\end{table}
\section{Takeaways}
\label{sec:takeaways}

In this section, we summarize the main takeaway messages that we determine from our discussion.

\begin{formal}
\textbf{Takeaway 1:} \textit{Closed model architectures mitigate several adversarial threats across all levels of \ac{av} automation.}
\end{formal}

As demonstrated by~\cite{alecci2023dumb}, access to the model architecture is the most critical aspect to consider when defending against adversarial attacks.
Although adhering to the ``security by obscurity'' paradigm, prevalent in the automotive industry, restricting access to these resources is the best solution until effective defenses against \ac{ml} attacks are developed.

\begin{formal}
\textbf{Takeaway 2:} \textit{A well-defined \ac{odd} is a cornerstone of \ac{av} security while progressing towards full automation.}
\end{formal}

As seen while increasing the level of automation in Section~\ref{sec:claims}, the definition of the \ac{odd} is fundamental for ensuring safety.
Setting boundaries on the automated tasks and formalizing the human intervention can mitigate risks while climbing towards the complete automation of \acp{av}.

\begin{formal}
\textbf{Takeaway 3:} \textit{Securing autonomous driving demands collective standards and innovation, considering the risks of a realistic threat model while prioritizing advantages over human driving.}
\end{formal}

Securing a trustworthy future demands collaboration among industries, academia, and regulators in the dynamic landscape of autonomous driving.
Establishing unified standards for security, transparency, and ethics is essential.
As such, the balance between innovation and resilience defines the path toward a secure autonomous landscape.
\section{Conclusions}
\label{sec:conclusions}

The literature surrounding \acp{av} highlights their crucial role in enhancing road safety by reducing collisions.
With precise environmental sensing, these vehicles offer the potential to revolutionize safety.
Additionally, studies suggest \acp{av} could contribute to a more sustainable transportation landscape by lowering fuel emissions.

\paragraph{\textit{Contribution}}
In this paper, we evaluated the risks of autonomous driving and established a pragmatic threat model highlighting the delicate balance between \ac{av} advantages and security challenges.
We explored whether autonomous driving systems can outperform human drivers under adversarial attacks, particularly considering evasion attacks.
Our vulnerability evaluation comprised all \ac{sae} automation levels in light of the realistic threat model.
Furthermore, we reviewed the \ac{av} attack literature and suggested security measures for \ac{ai} implementation in \acp{av}.
Our contributions aim to advance discourse and technological development while ensuring the safe integration of \ac{ai} into safety-critical operations.

\paragraph{\textit{Future Works}}
In the ongoing pursuit of securing autonomous driving, future research should refine threat models and innovate defense mechanisms.
A crucial direction is assessing the adaptability of \ac{ai} systems to emerging adversarial strategies and improving the resilience of both human and \ac{ai} drivers against evolving threats.
Considering the diverse levels of \ac{sae} automation in \acp{av}, developing targeted countermeasures becomes imperative for a more robust autonomous driving landscape.
Furthermore, an essential future work direction involves evaluating the criteria identified on a practical testbed.
This testbed could simulate real-world conditions, allowing for the empirical validation of the criteria's efficacy in assessing the security of autonomous driving systems against adversarial challenges.
Addressing feasibility and practicality remains central to advancing the field.

\balance
\bibliographystyle{plain}
\bibliography{references}

\end{document}